\documentclass[aps,prl,10pt,twocolumn,superscriptaddress,floatfix,nofootinbib,  showpacs,longbibliography]{revtex4-2}

\usepackage{graphicx}
\usepackage{braket}
\usepackage[utf8]{inputenc}  
\usepackage[T1]{fontenc}    
\usepackage[british]{babel}  
\usepackage[sc,osf]{mathpazo}

\usepackage{amsmath,amssymb,amsthm}
\usepackage{xcolor}
\usepackage{algorithm}
\usepackage{algorithmic}
\usepackage{natbib}
\usepackage{lineno}
\usepackage[normalem]{ulem} 
\usepackage{adjustbox} 
\usepackage{tikz}
\usetikzlibrary{quantikz2}
\usepackage[colorlinks=true,linkcolor=red,citecolor=blue,urlcolor=blue]{hyperref}
\usepackage{comment}

\definecolor{darkgreen}{rgb}{0.0, 0.5, 0.0}

\DeclareMathOperator{\Tr}{Tr}

\usepackage[colorinlistoftodos,prependcaption,textsize=tiny]{todonotes}
\newcommandx{\cmc}[2][1=]{\todo[linecolor=red,backgroundcolor=red!10,bordercolor=red,#1]{#2}}
\newcommandx{\san}[2][1=]{\todo[linecolor=blue,backgroundcolor=blue!10,bordercolor=blue,#1]{#2}}

\begin{document}

\title{Photonic Simulation of Beyond-Quantum Nonlocal Correlations (e.g. Popescu-Rohrlich Box) with Non-Signaling Quantum Resources}

\author{Kunal Shukla}
\affiliation{Quantum Optics \& Quantum Information Laboratory, Dept. of Electronic Systems Engineering, Indian Institute of Science, Bengaluru 560012, India}
\affiliation{Dept. of Instrumentation \& Applied Physics, Indian Institute of Science, Bengaluru 560012, India}

\author{Anirudh Verma}
\affiliation{Quantum Optics \& Quantum Information Laboratory, Dept. of Electronic Systems Engineering, Indian Institute of Science, Bengaluru 560012, India}

\author{Kanad Sengupta}
\affiliation{Quantum Optics \& Quantum Information Laboratory, Dept. of Electronic Systems Engineering, Indian Institute of Science, Bengaluru 560012, India}
\affiliation{Dept. of Instrumentation \& Applied Physics, Indian Institute of Science, Bengaluru 560012, India}

\author{Sanchari Chakraborti}
\affiliation{Quantum Optics \& Quantum Information Laboratory, Dept. of Electronic Systems Engineering, Indian Institute of Science, Bengaluru 560012, India}

\author{Manik Banik}
\affiliation{Department of Physics of Complex Systems, S. N. Bose National Center for Basic Sciences,
Block JD, Sector III, Salt Lake, Kolkata 700106, India}

\author{C. M. Chandrashekar}
\affiliation{Quantum Optics \& Quantum Information Laboratory, Dept. of Electronic Systems Engineering, Indian Institute of Science, Bengaluru 560012, India}

\begin{abstract}

Bell nonlocality exemplifies the most profound departure of quantum theory from classical realism. Yet, the extent of nonlocality in quantum theory is intrinsically bounded, falling short of the correlations permitted by the relativistic causality (the no-signaling) principle. A paradigmatic example is the Popescu–Rohrlich correlation: two distant parties sharing arbitrary entanglement cannot achieve this correlation, though it can be simulated with classical communication between them. Here we show how such post-quantum correlations can instead be simulated using intrinsically non-signaling physical resources, and implement the proposed scheme using a quantum circuit on a four-qubit photonic platform. Unlike the conventional approaches, our method exploits dynamical correlations between distinct physical systems, with intrinsic randomness suppressing any signaling capacity. This enables the realization of post-quantum correlations both with and without entanglement. We also analyze how the simulation scheme extends to beyond quantum nonlocal correlations in multipartite systems. Our experimental demonstration using a photonic system establishes a versatile framework for exploring post-quantum correlations in both foundational settings and as a resource for computation and security applications.

\end{abstract}

\maketitle
\noindent
\textit{Introduction.--} Quantum mechanics, the most successful theory to describe the microscopic phenomena, leads to predictions that defy classical worldviews. For instance, Bell’s seminal 1964 theorem \cite{Bell1964} establishes that local measurement outcomes on composite quantum systems cannot always be reproduced by local-causal models \cite{Bell1966,Mermin1993,Brunner2014}. This prediction has been confirmed in a series of landmark experiments through violations of Bell inequalities \cite{Freedman1972,Aspect1981,Aspect1982(1),Aspect1982(2),Zukowski1993,Weihs1998,Tittel1998,Hensen2015,Handsteiner2017,BIG2018}. One of the most studied inequalities is the Clauser–Horne–Shimony–Holt (CHSH) inequality \cite{Clauser1969}, where classical correlations are bounded by $2$, but quantum mechanics allows violations up to Tsirelson’s bound of $2\sqrt{2}$ \cite{Cirelson1980}. Popescu and Rohrlich, however, showed that nonlocality can be of stronger form: their hypothetical PR-box achieves the algebraic maximum $4$ of the CHSH expression, while respecting the no-signaling (NS) principle thereby prohibiting faster than light communication \cite{Popescu1994} (see also \cite{Popescu2014}). Despite satisfying NS, the PR correlation carries striking consequences. For instance, they trivialize communication complexity, allowing any function to be computed with only a constant amount of communication between two parties \cite{vanDam2012}. This apparent absurdity has led to the conjecture that nontrivial communication complexity is a fundamental axiom of quantum mechanics \cite{Brassard2006,Buhrman2010}. Subsequently, several other principles have been identified that constrain the set of physically realizable correlations \cite{Pawowski2009,Navascus2009,Fritz2013,Cabello2013,Naik2022,Patra2023}.

In quantum theory, nonlocality typically arises from entangled states shared between spatially separated parties. In such a spacelike scenario, measurements on the local parts of the entangled states produce nonlocal correlations bounded by Tsirelson's limit. In contrast, timelike scenarios—where classical communication is permitted between the parties—allow simulation of arbitrary nonlocal correlations, including the PR correlations \cite{Hall2011,Toner2003,Brassard2016}. These two regimes can be unified in the quantum-circuit picture using entangling gates that induce global interaction among distinct quantum systems. However, such a gate can serve either as a non-signaling resource or as a signaling resource depending on how its input preparations are initialized \cite{Bennett2003,Harrow2004,Harrow2010,Hayden2020}.

Here, we show that restricted access to multi-qubit gates can be used to simulate PR correlation, even when they  act as non-signaling resources. Crucially, these beyond-quantum correlations are not generated through the conventional mechanism of creating entanglement between distinct quantum systems. Instead, the gate dynamically controls correlations across different physical systems, while intrinsic randomness within the gate ensures that the overall process remains non-signaling. We implement this mechanism on a four-qubit photonic platform with deterministic quantum gate operations using polarization-path degrees of freedom of photons \cite{Maya2024, Kanad2025}, validating the recently proposed one-time-pad (OTP) model for PR correlations \cite{Sidhardh2024}, which can be viewed as a De-Broglie-Bohm type model for PR correlation \cite{Bub2013}. In our experiment, the suitable combination of Hadamard, CNOT, and Toffoli gates suppresses the signaling capacity of the resource. We realize a photonic circuit that generates a two-qubit maximally entangled state with fidelity $93.9\%$, simulates CHSH violations exceeding $3.9$—well beyond Tsirelson's bound—and with visibilities above $99\%$ across multiple bases. In addition, numerical simulations confirm that PR correlations can be obtained without entanglement, but through the preparation of classically correlated states following the same underlying principle of harnessing dynamic correlations and randomness. We further extend our framework to multipartite settings, demonstrating the potential of quantum circuits as versatile platforms for simulating beyond-quantum resources.\\

\noindent
\textit{Theory.--} A typical bipartite Bell scenario involves two spatially separated parties, Alice and Bob, who after receiving classical inputs $x\in\mathcal{X}$ and $y\in\mathcal{Y}$, respectively, from a referee returns classical outputs $a\in\mathcal{A}$ and $b\in\mathcal{B}$. Throughout, we assume the input and output sets to be of finite cardinalities. Repeated runs of the experiment generate an input–output correlation (also called a behavior) $P\equiv\{p(a,b|x,y)~|~\sum_{a,b}p(a,b|x,y)=1~\forall~x,y\}$. A correlation is called non-signaling (NS) if neither party can signal to the other, i.e.,
\begin{subequations}
\begin{align}
P(a|x,y) &= P(a|x,y'),~~\forall~~ a,x,y,y', \label{eq:ns_Pa}\\
P(b|x,y) &= P(b|x',y),~~ \forall~~ b,x,x',y.
\label{eq:ns_Pb}
\end{align}
\end{subequations}
A correlation is called quantum if it can be obtained through local measurements on a bipartite quantum system, namely if it admits a realization of the form 
\begin{equation}
   p(a,b|x,y)=\mathrm{Tr}[(\pi^a_x\otimes\pi^b_y)\rho_{AB}], 
\end{equation}
for some $\rho_{AB}\in\mathcal{D}(\mathcal{H}_A\otimes\mathcal{H}_B)$, where $\mathcal{H}_A$ and $\mathcal{H}_B$ denote the Hilbert spaces corresponding to Alice and Bob, respectively. Moreover, the operators ${\pi^a_x}\subseteq \mathcal{P}(\mathcal{H}_A)$ and ${\pi^b_y}\subseteq \mathcal{P}(\mathcal{H}_B)$ respectively satisfy $\sum_a \pi^a_x=\mathbf{I}_A$ and $\sum_b \pi^b_y=\mathbf{I}_B$, for all $x, y$. Here, $\mathcal{D}(\cdot)$ and $\mathcal{P}(\cdot)$ denote the sets of density operators on $\mathcal{H}_A \otimes \mathcal{H}_B$ and positive operators on $\mathcal{H}_A$ or $\mathcal{H}_B$, respectively, while $\mathbf{I}$ denotes the identity operator. For a finer classification of quantum correlations, we refer to the recent breakthrough result of Slofstra \cite{Slofstra2019}. A correlation is called classical (more specifically Bell local) if it admits a local-causal decomposition, i.e.,
\begin{align}
p(a,b|x,y)=\int_{\Lambda}d\lambda\,\mu(\lambda)\,p(a|x,\lambda)p(b|y,\lambda),    
\end{align}
where $\mu(\lambda)$ is a probability distribution over hidden variables $\lambda\in\Lambda$ \cite{Brunner2014}. The sets $\mathcal{N},~\mathcal{Q},~\mathcal{L}$ of NS, quantum, and local correlations, respectively, satisfy the strict set inclusions $\mathcal{L}\subsetneq\mathcal{Q}\subsetneq\mathcal{N}$. In the simplest binary-input–binary-output scenario the CHSH inequalities turns out to be the necessary and sufficient criteria to establish nonlocality of a correlation \cite{Fine1982}. Denoting $\mathcal{A}=\mathcal{B}=\{+1,-1\}$, the CHSH inequalities, up to local relabelings, take the form
\begin{equation}
\mathcal{S}:=|\braket{a_0b_0}+ \braket{a_0b_1}+\braket{a_1b_0}-\braket{a_1b_1}|\le2, 
\end{equation}
where $\langle a_x b_y\rangle:=\sum_{a,b=\pm1}ab \; p(a,b|x,y)$. 

The quantum set $\mathcal{Q}$ is highly nontrivial and lies strictly between $\mathcal{L}$ and $\mathcal{N}$ \cite{Barizien2025}. Local correlations achieve a maximum CHSH score of $2$, while quantum correlations can reach the Tsirelson's bound of $2\sqrt{2}$ \cite{Cirelson1980}, achievable by a maximally entangled two-qubit state with suitable measurements. By contrast, the Popescu–Rohrlich (PR) box correlation $P_{PR}$ \cite{Popescu1994}, defined by
\begin{align}
p(a,b|x,y)=
\begin{cases}
\tfrac{1}{2}, & \text{if } ab=(-1)^{xy},\\
0, & \text{otherwise},
\end{cases}\label{pr}    
\end{align}
achieves the algebraic maximum $4$ of CHSH expression.
\begin{figure}[t!]
\centering
\includegraphics[width=0.95\linewidth]{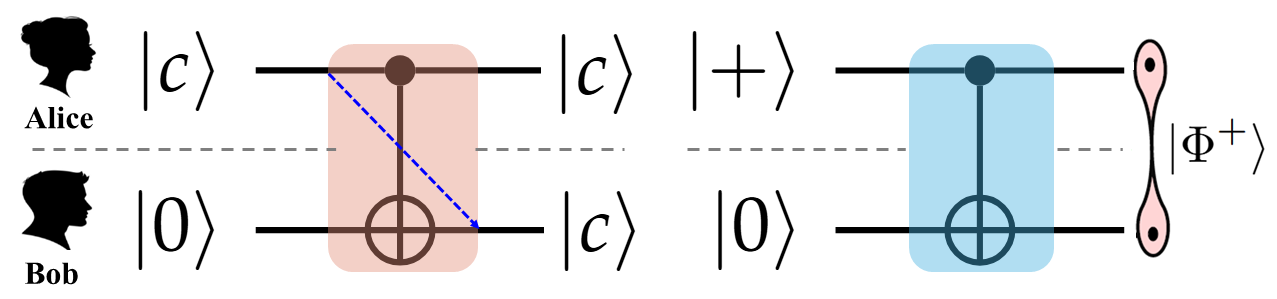}
\caption{[Left]: A two-qubit $\mathrm{CNOT}$ gate enables one-bit classical communication from Alice to Bob (a time-like signaling resource) when Alice is free to prepare her qubit in $\ket{c}\in\{\ket{0},\ket{1}\}$. [Right]: If Alice is restricted to initializing her qubit only in $\ket{+}$, the gate instead acts as a space-like no-signaling resource, generating a maximally entangled two-qubit state $\ket{\Phi^+}$ shared between Alice and Bob.}
\label{fig:CNOT}
\vspace{-.3cm}
\end{figure}

\noindent
{\it Nonlocal capacity of multi-qubit gates.--} Multi-qubit gates, such as the $\mathrm{CNOT}$, $\mathrm{SWAP}$, and $\mathrm{Toffoli}~(\mathrm{\mathrm{CCNOT}})$ gates \cite{Cirac1995,Monroe1995,Anderlini2007,Lanyon2008}, play a central role in quantum computing architectures \cite{Kitaev1997,Dawson2006}. Such gates can be regarded as oracles, wherein different parties input their respective quantum systems, prepared in suitable states, and receive a transformed state as output. The gate mediates a global interaction among distinct systems, effectively serving as a nonlocal resource \cite{Bennett2003,Harrow2004,Harrow2010,Hayden2020}. For example, a two-qubit $\mathrm{CNOT}$ gate with Alice’s qubit as the control and Bob’s qubit as the target enables one-bit classical communication from Alice to Bob: the target qubit is prepared in $\ket{0}$, while Alice encodes the bit value she wishes to send by preparing the control qubit in either $\ket{0}$ (to send `$0$') or $\ket{1}$ (to send `$1$') as can be seen in Fig.~\ref{fig:CNOT}~[Left]). On the other hand, if Alice’s access to the $\mathrm{CNOT}$ oracle is restricted — so that she can only initialize her qubit in $\ket{+} := (\ket{0}+\ket{1})/${\footnotesize $\sqrt{2}$} (see Fig.~\ref{fig:CNOT}~[Right]) — then no classical communication from Alice to Bob is possible. Nevertheless, this restricted access still enables them to prepare a maximally entangled state $\ket{\Phi^+} :=(\ket{00}+\ket{11})/${\footnotesize $\sqrt{2}$}, which in turn can be used to achieve the CHSH value $2\sqrt{2}$. \\

\begin{figure}[t!]
\centering
\adjustbox{width=.95\columnwidth}{
\includegraphics{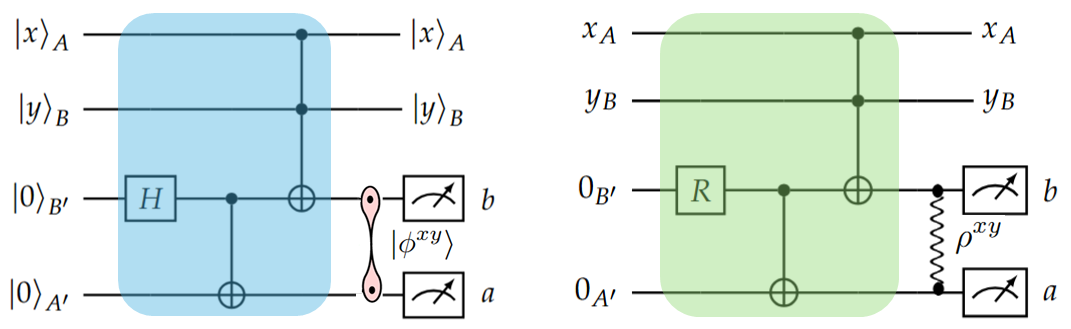}}
\caption{[Left]: Four-qubit quantum oracle (blue shaded box) -- qubits $A,A'$ correspond to Alice’s inputs and $B,B'$ to Bob’s. When the primed qubits are initialized in $\ket{0}$ and $A$ and $B$ are respectively prepared in $\ket{x}$ and $\ket{y}$ with $x,y\in \{0,1\}$, the output state of the primed systems after the oracle action becomes $\ket{\phi^{xy}}$ (see Eq.~\ref{qpr}). [Right]: Corresponding classical circuit, obtained by replacing qubits and quantum gates with their classical analogues. In this case, after the action of the classical oracle (green shaded box), the state of the primed systems becomes $\rho^{xy}_{B'A'}=\tfrac{1}{2}\big[(x y)_{B'}(0)_{A'}+(x y\oplus1)_{B'}(1)_{A'}\big]$.}
\label{fig:PR-Simulation}
\vspace{-.3cm}
\end{figure}

\noindent{\it Simulating beyond-quantum nonlocal correlation.--} Here, we show how the PR correlation can be simulated using a non-signaling quantum oracle. Consider the four-qubit oracle as shown in Fig.~\ref{fig:PR-Simulation}~[Left]. Systems $A$ and $A'$ are held by Alice, while $B$ and $B'$ are with Bob. The primed systems are always initialized in ket $\ket{0}$, whereas, Alice and Bob can prepare their respective unprimed systems only in computational basis states. After the action of a Hadamard and a $\mathrm{CNOT}$ gate, the state of the primed systems becomes $\ket{\Phi^{+}}_{B'A'}$. Subsequently, the $\mathrm{Toffoli}$ gate, with the unprimed systems $A$ and $B$ as controls and $B'$ as the target, transforms the primed systems into the state
\begin{align}
\ket{\phi^{xy}}_{B'A'}:=\tfrac{1}{\sqrt{2}}\big(\ket{x y}\ket{0}+\ket{x y\oplus1}\ket{1}\big)_{B'A'} ~,\label{qpr}
\end{align}
where $\oplus$ denotes addition modulo $2$. If Alice and Bob then measure their respective primed qubits in the computational basis, this oracle simulates the PR correlation (Eq.~\ref{pr}). At this point we note that our oracle mimics the recently proposed classical-to-quantum non-signaling boxes \cite{Ferrera2024}.

It is instructive to analyze this simulation protocol in more detail. First, with the restricted input preparations, the oracle does not enable communication between the parties and thus acts as a non-signaling resource. While the $\mathrm{Toffoli}$ gate, in principle, allows one bit of communication from Alice's unprimed system to Bob's primed system, the prior Hadamard and $\mathrm{CNOT}$ operations entangle $A'$ with $B'$, thereby distributing the potential signaling effect across their correlations. As a result, the oracle as a whole remains non-signaling (see Supplementary Material). In this sense, our protocol is reminiscent of the recently proposed OTP model for nonlocal correlations \cite{Sidhardh2024}, where nonlocality is interpreted as signaling at the hidden-variable level, but the randomness in the hidden-variable distribution ensures compatibility with the operational no-signaling condition. In our case, the simulation protocol dynamically controls the correlations between the primed qubits.

Notably, entanglement is not essential for simulating the PR correlation. Replacement of the quantum oracle with its classical analogue (see Fig.~\ref{fig:PR-Simulation}:~[Right]), where the Hadamard gate is substituted by a randomizing gate $\mathrm{R}(\alpha)=\tfrac{1}{2}\alpha+\tfrac{1}{2}\bar{\alpha},~\forall~\alpha\in\{0,1\}$, reproduces PR correlations exactly . However, entanglement enables the simulation protocol more robust, as it allows PR correlations across multiple measurement bases, while the classically correlated state allows PR correlations only in computational bases (see Supplementary Material).

Figure~\ref{fig: CHSH_values}(a) presents our experimental results establishing that the dynamically controlled entangled state $\ket{\phi^{xy}}$ [given in Eq.~\ref{qpr}] simulates the PR correlation across multiple measurement bases. The state $\ket{\phi^{xy}}$ is prepared through the four-qubit photonic setup depicted in Fig.~\ref{fig: Expt_schematic}. CHSH values are measured for three different local measurement bases: computational $\{\ket{H}, \ket{V}\}$, diagonal $\{\ket{D}, \ket{A}\}$, and circular $\{\ket{L}, \ket{R}\}$ bases and the corresponding results are plotted in Fig.~\ref{fig: CHSH_values}(a). The PR correlation is obtained only in the computational and circular settings, where violations surpass the Tsirelson's bound and approach $\mathcal{S}=4$; the diagonal basis remains limited to $\mathcal{S}=2$, consistent with the theory (see Supplementary Material). Figure~\ref{fig: CHSH_values}(b) further shows how the CHSH values for the entangled state $\ket{\phi^{xy}}$ vary with the polar angle \(\theta\) associated with the local measurement basis states \(\{\ket{\psi}:=\cos(\theta/2)\ket{0}+e^{i\phi}\sin(\theta/2)\ket{1},\ket{\psi^\perp}\}\). Here we sweep \(\theta\) through the X–Z plane (i.e., \(\phi=0\)) of the Bloch sphere and find that the experimentally obtained CHSH values agree with the theoretical prediction (see the Supplementary Material). Next, we turn to the unentangled mixed state $\rho^{xy}$: numerical simulations of $\rho^{xy}$ (Fig.~\ref{fig: CHSH_values}(c)) confirm PR correlations obtained without entanglement in the computational measurement  basis. We further observe Tsirelson's bound violations for other bases, with the CHSH value being independent of the azimuthal angle ($\phi$).\\


\begin{figure*}[t!]
\centering
\adjustbox{width=2\columnwidth}{%
\includegraphics{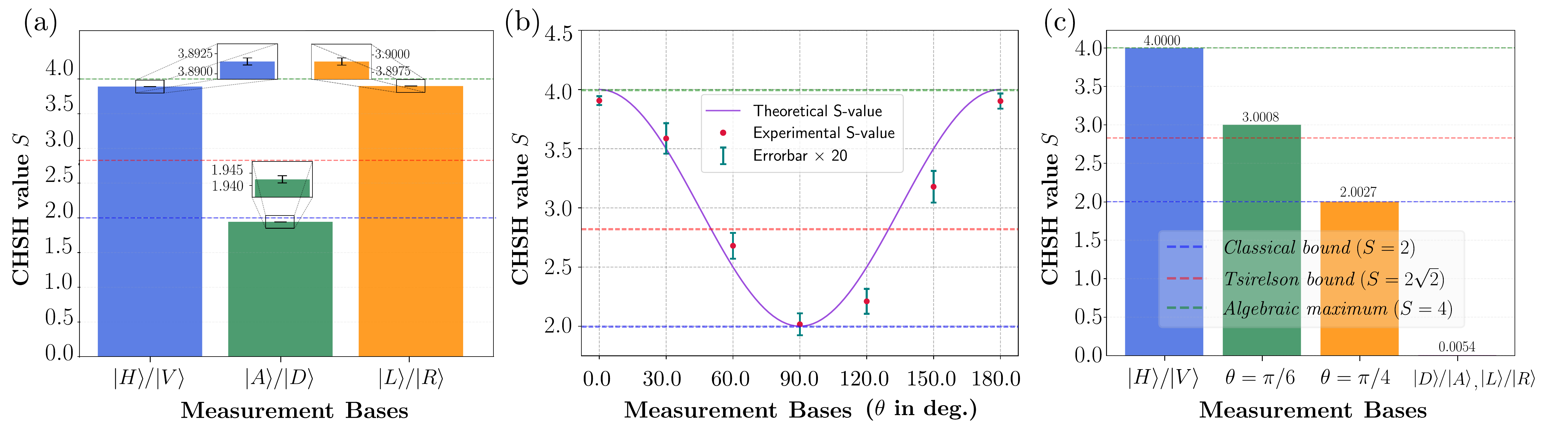}}
 \caption{(a) Experimentally obtained CHSH values \(\mathcal{S}\) when Alice and Bob perform local measurements on the entangled state \(\ket{\phi^{xy}}\) in the bases listed along the x-axis. Computational basis measurement \(\{\ket{H},\ket{V}\}\) and circular basis measurement \(\{\ket{L},\ket{R}\}\) approach the algebraic maximum \(\mathcal{S}=4\) (PR correlations), whereas diagonal basis \(\{\ket{A},\ket{D}\}\) attains at most \(\mathcal{S}=2\). (b) CHSH values for local measurement bases \(\{\ket{\psi}:=\cos(\theta/2)\ket{0}+e^{i\phi}\sin(\theta/2)\ket{1},\ket{\psi^\perp}\}\) parameterized by the polar angle \(\theta\),  with \(\phi=0\) within the Bloch sphere. Experimentally obtained CHSH values (red dots) for the entangled state \(\ket{\phi^{xy}}\) agree with the theoretical expectations (solid violet curve) within the experimental errors (teal bars). (c) Numerical simulation of CHSH values for the unentangled mixed state \(\rho^{xy}\). The measurement bases are parameterized by \(\theta\) as above; the final bar confirms that, unlike the entangled state \(\ket{\phi^{xy}}\), the CHSH value \(\mathcal{S}\) for \(\rho^{xy}\) is independent of the azimuthal parameter \(\phi\).}
 \label{fig: CHSH_values}
\end{figure*}

\begin{figure*}[t!]
\centering
\adjustbox{width=1.90\columnwidth}{%
\includegraphics{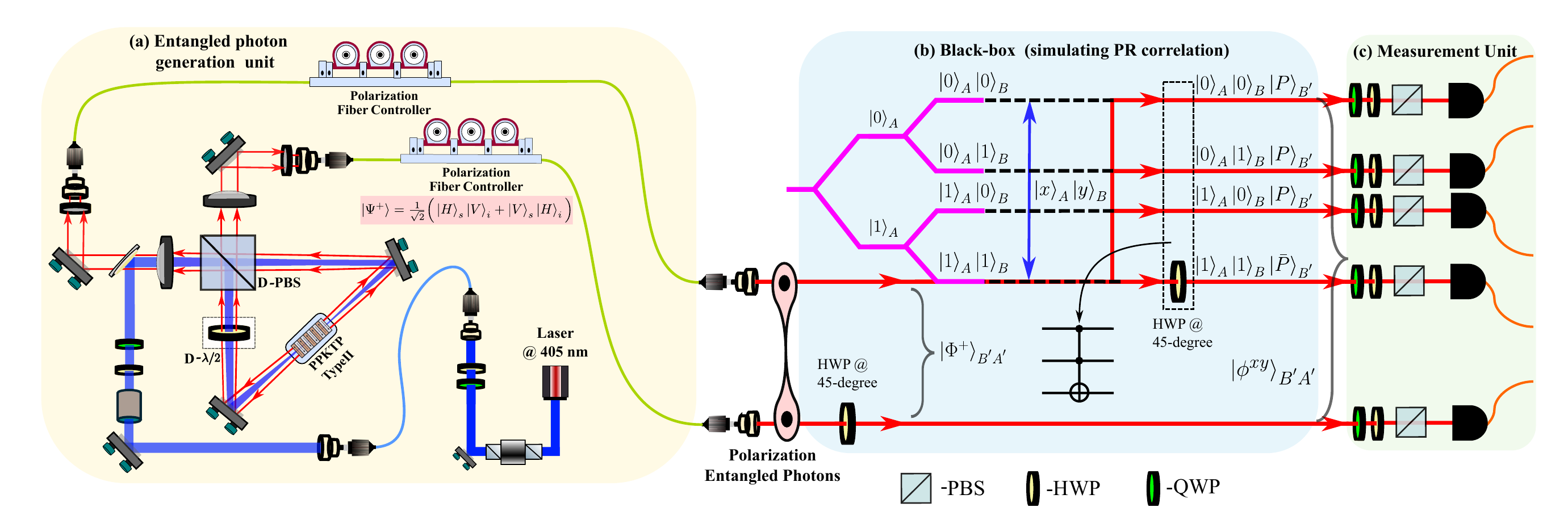}}
\caption{Experimental setup. (a) State Preparation: A polarization-entangled state $\ket{\Psi^{+}}_{B'A'} = \left(\ket{H}_{B'}\ket{V}_{A'} + \ket{V}_{B'}\ket{H}_{A'}\right)/\footnotesize{\sqrt{2}}$ is prepared through SPDC process using a type-II PPKTP crystal in a Sagnac interferometer configuration. The photon pairs are then coupled to two single mode fibers and two polarization fiber controllers are used to maintain entanglement during the fiber transmission. (b) The oracle: The state $\ket{\Psi^{+}}_{B'A'}$ is converted to $\ket{\Phi^{+}}_{B'A'}$ using a half-wave plate (HWP) as $\sigma_{x}$ operator in the path of one of the photons. A photon from the polarization-entangled pair is then sent through one of the four distinct paths represented by $\ket{x}_{A}\ket{y}_{B}$, with $x,y \in \{0,1\}$. Followed by the path selection, a Toffoli gate is applied to the path-path-polarization encoded state with the path d.o.f. in $A$ and $B$ as the controls and polarization d.o.f. in $B'$ as the target, resulting in the state $\ket{x}_{A}\ket{y}_{B}\ket{\phi^{xy}}_{B'A'}$. (c) Measurement unit: The entangled state $\ket{\phi^{xy}}_{B'A'}$ is measured in different polarization basis using QWP, HWP and PBS combinations, and recording the coincidences between the photon in one of the four paths and the other photon.}

\label{fig: Expt_schematic}
\end{figure*}

\noindent
\textit{Experimental implementation. --} We implement the proposed four-qubit scheme on a photonic platform using polarization-entangled photon pairs generated via type-II SPDC in a PPKTP crystal within a Sagnac interferometer. The source produces the Bell state $\ket{\Psi^{+}}_{B'A'} = \big(\ket{H}_{B'}\ket{V}_{A'} + \ket{V}_{B'}\ket{H}_{A'}\big)/\footnotesize{\sqrt{2}}$, with a measured CHSH violation of $\mathcal{S} = 2.708 \pm 0.02$ (see Supplementary Material for details). In our model for simulating PR correlations, this state serves as the resource for inputs $(x,y)=(1,1)$ [as can be seen from Eq.~\ref{qpr}]. For the other inputs $(0,0),(0,1),(1,0)$, the relevant resource is $\ket{\Phi^{+}}_{B'A'} =\big(\ket{H}_{B'}\ket{H}_{A'}+\ket{V}_{B'}\ket{V}_{A'}\big)/\footnotesize{\sqrt{2}}$, which we obtain from $\ket{\Psi^{+}}_{B'A'}$ by applying $\sigma_{x}$ operation to one of the polarization qubits (here, on $A'$). The $\sigma_{x}$ operator is experimentally realized with a half-wave plate (HWP) with its fast axis oriented at $45^\circ$. A schematic of the entangled photon source is shown in Fig.~\ref{fig: Expt_schematic}(a).

A four-qubit state with path-path-polarization encoding for the signal photon and polarization encoding for the idler photon is prepared in an experimental setting \cite{Maya2024, Kanad2025} for the simulation of the PR correlation. A general path–path–polarization encoding can be obtained using three interferometers with appropriate relative phase tuning. The corresponding Hilbert space is $\mathcal{H}_{\text{total}} \equiv (\mathcal{H}_{p_1}\otimes \mathcal{H}_{p_2}\otimes \mathcal{H}_{\text{pol}})_{S}\otimes (\mathcal{H}_{\text{pol}})_{I}$, where $p_1$ and $p_2$ represent the path degree of freedom. Here, the signal photon is coherently distributed among four path modes $\ket{0}\ket{0}, \ket{0}\ket{1}, \ket{1}\ket{0}, \ket{1}\ket{1} \in \mathcal{H}_{p_1}\otimes \mathcal{H}_{p_2}$, representing the states $\ket{x}_{A}\ket{y}_{B}$ as shown in Fig.~\ref{fig: Expt_schematic}(b). The interferometric phase settings determine the classical inputs $(x,y)$, while the third qubit is given by the polarization d.o.f. of the signal photon. To generate the resource state $\ket{\phi^{xy}}_{B'A'}$, a Toffoli gate is implemented with the two path qubits as controls and the polarization qubit with Bob as target. Experimentally, the Toffoli operation is realized by inserting a HWP at $45^\circ$ in the $\ket{1}\ket{1}$ path, which flips the polarization $\ket{H}\leftrightarrow\ket{V}$, thereby converting $\ket{\Phi^{+}}_{B'A'}$ into $\ket{\Psi^{+}}_{B'A'}$. Temporal indistinguishability of the photons coming from the other paths is ensured with the use of identical thickness glass plates. The resulting oracle mimics the Toffoli gate in the path–polarization encoding, with input state $\ket{x}_{A}\ket{y}_{B}\otimes (\ket{H}_{B'}\ket{H}_{A'}+\ket{V}_{B'}\ket{V}_{A'})/\footnotesize{\sqrt{2}}$ [Fig.~\ref{fig: Expt_schematic}(b)].

In experiment, the CHSH parameter is evaluated using the expression $ \mathcal{S} = \left| \sum_{x,y \in \{0,1\}} (-1)^{x \cdot y} \, E(a_{x}, b_{y}) \right|$, where $E(a_x,b_y)$ is calculated from coincidence counts $C(a^\circ,b^\circ|x,y)$ using the formula
\begin{align}
E(a_x,b_y) = \frac{\sum_{a^\circ,b^\circ} (-1)^{1+\delta_{a^\circ b^\circ}} C(a^\circ,b^\circ|x,y)}{\sum_{a^\circ,b^\circ} C(a^\circ,b^\circ|x,y)},
\end{align}
here $a^\circ,b^\circ\in\{0^\circ,45^\circ\}$ denote the half-wave plate angles before polarizing beam splitter (PBS). Under ideal conditions the expected value is $\mathcal{S}=4$; in our setup we obtain $\mathcal{S} = 3.91 \pm 0.002$, close to the maximum achievable value.

\noindent\textit{Conclusion. --} For the first time, we demonstrate that beyond-quantum nonlocal correlations can be simulated by embedding dynamic correlations into quantum circuits functioning as no-signaling oracles. Multi-qubit gates such as CNOT and Toffoli, together with a Hadamard gate and restricted input access, generate correlations that adapt to the parties’ inputs and reproduce PR-box behavior. Our four-qubit photonic experiment confirms this prediction, yielding CHSH values approaching the algebraic maximum of $4$, far beyond the Tsirelson bound. Notably, the protocol also functions with a classical oracle, in which case entanglement is not required. In the quantum setting, however, entanglement enhances robustness and expands the range of local measurement bases that sustain beyond-quantum correlations. Extensions of the protocol to multipartite beyond-quantum nonlocality are presented in the Supplementary Material.

From a foundational perspective, our results provide experimental support for the recently proposed one-time-pad model of PR correlation \cite{Sidhardh2024} as well as for De Broglie–Bohm type construction underlying PR correlation \cite{Bub2013}. Practically, they demonstrate that multi-qubit gate oracles, even when constrained to respect no-signaling, can simulate resources valuable for communication complexity \cite{vanDam2012}. Looking ahead, it will be interesting to explore whether device-independent protocols, such as secure key distribution and random-number generation \cite{Pironio2010,Colbeck2011,Colbeck2012,Vazirani2014}, can benefit from such dynamic oracles. A central open question is whether Bell-type inequalities can be formulated whose beyond-quantum violations cannot be simulated by dynamically correlated states. Resolving this would deepen our foundational understanding of nonlocality and inspire new protocols in quantum communication, cryptography and quantum algorithms.

\begin{acknowledgements}
\noindent{\bf Acknowledgement}: We acknowledge funding support from the National Quantum Mission, an initiative of the Department of Science and Technology, Govt. of India.
\end{acknowledgements}

\bibliography{OTP_Pad2}

\clearpage
\onecolumngrid

\part*{Supplemental material for ``Photonic Simulation of Beyond-Quantum Nonlocal Correlations (e.g. Popescu-Rohrlich Box) with Non-Signaling Quantum Resources''}

\setcounter{figure}{0}
\setcounter{table}{0}
\setcounter{equation}{0}
\renewcommand{\thefigure}{S\arabic{figure}}
\renewcommand{\thetable}{S\arabic{table}}
\renewcommand{\theequation}{S\arabic{equation}}

\section{A. Simulating beyond-quantum nonlocality and no-signaling property of the oracles}
In the main article, we claimed that the states $\ket{\phi^{xy}}_{B'A'}$ and $\rho^{xy}_{B'A'}$ with and without entanglement, respectively, generated by circuits given in Fig.~2 of the main text, are no-signaling resources which can produce nonlocal behaviors. Here, we discuss detailed proofs of those claims. We begin with mathematically formalizing CHSH scenario in our implementation. 

In bipartite Bell scenario, the joint system of Alice and Bob is described by the Hilbert space $\mathcal{H}=\mathcal{H}_A \otimes \mathcal{H}_B$ where both $\mathcal{H}_A$, and $\mathcal{H}_B$ is two dimensional. In our case, they can share an entangled state $\ket{\phi^{xy}}_{B'A'}$ or equivalently $\ket{\phi^{xy}}_{A'B'}$ (where we have changed the order in the tensor product keeping Alice's state first) or an unentangled mixed state $\rho^{xy}_{B'A'} \rightarrow \rho^{xy}_{A'B'}$. Here, we represent states $\ket{\phi^{xy}}_{A'B'}$ and $\rho^{xy}_{A'B'}$ by density matrices $\rho^*_q$ and $\rho^\star_c$, respectively. The shared state $\rho^\star (\rho^*_q \text{ or } \rho^\star_c)$ is a positive Hermitian linear operator of unit trace acting on $\mathcal{H}$. The \emph{process} Alice and Bob use in every round to produce outputs $a,b \in \mathcal{R}=\{-1, 1\}$ after receiving inputs $x,y \in \mathcal{I}= \{0,1\}$ is the following. Based on their inputs, they choose measurement bases described by the sets $\mathcal{M}^x_A = \{\Pi^x_a|a\in\mathcal{R}\}$ for Alice and $\mathcal{M}^y_B = \{\Pi^y_b|b\in\mathcal{R}\}$ for Bob, where $\Pi^x_a$ and $\Pi^y_b$ are projection-valued measures (PVMs) acting on $\mathcal{H}_A$ and $\mathcal{H}_B$, respectively. Consequently, for each $x$ and $y$, we have a set of PVMs $\mathcal{M}^x_A$ and $\mathcal{M}^y_B$, respectively. After choosing the measurement bases, they perform local measurements on the shared state $\rho^\star$ and get outcomes $a$ and $b$. The behavior $\mathcal{P}^\star$ given by such processes is 

\begin{equation}
	\mathcal{P}^\star :\quad P(a,b\mid x,y) = \Tr(\rho^\star \;\Pi^x_a\otimes \Pi^y_b),
\end{equation}
where $\rho^\star$ could be $\rho^\star_q$ or $\rho^\star_c$ depending on whether entangled or unentangled state is used. In the following, we want to systematically establish:
Under appropriate choice of $\mathcal{M}^x_A$ and $\mathcal{M}^y_B$, behaviors shown by the states $\rho^\star_q$ and $\rho_c^\star$ are PR behaviors, i.e., $\mathcal{P^\star}= \mathcal{P}_{PR}$.

We must first prove the no-signaling property of our states. We need to show that for any choice of measurement bases, $P(a \mid x,y) = P(a\mid x,y')$ and $P(b \mid x,y) = P(b\mid x',y)$ holds for all inputs and outputs. Consequently, we need to find an expression for $P(a,b\mid x,y) = \Tr(\rho^\star \;\Pi^x_a\otimes \Pi^y_b)$ for arbitrary measurement bases. PVMs $\Pi^x_a$ and $\Pi^y_b$ can be labeled by directions $\hat{a}_x , \hat{b}_y$ in the Bloch sphere $\mathbb{S}^2$, respectively. In more precise terms, 
\begin{equation}
	\Pi^x_a = \frac{1}{2}(\mathbb{I} +a \;\hat{a}_x\cdot\hat{\sigma}),
	\label{eq: Alice_measurement}
\end{equation}
where Alice maps input $x$ to direction $\hat{a}_x \in \mathbb{S}^2$, thereby characterizing the set $\mathcal{M}^x_A = \{\Pi^x_{+1}, \Pi^x_{-1}\}$. Similarly, Bob does the following mapping $y \mapsto \hat{b}_y \mapsto \mathcal{M}^y_B = \{\Pi^y_b|b\in\mathcal{R}\}$, where
\begin{equation}
	\Pi^y_b = \frac{1}{2}(\mathbb{I} +b \;\hat{b}_y\cdot\hat{\sigma}).
	\label{eq: Bob_measurement}
\end{equation}
In both the Eqs.~(\ref{eq: Alice_measurement}) and (\ref{eq: Bob_measurement}), ordered list $\hat{\sigma} = (\sigma_x, \sigma_y, \sigma_z)$ of Pauli operators multiply with the three components of the vectors $\hat{a}_x = (a^x_1, a^x_2, a^x_3)$ and $\hat{b}_y = (b^y_1, b^y_2, b^y_3)$. We are now ready to calculate the probabilities $P(a,b \mid x,y)$; let's begin with the entangled state: 

\begin{align*}
	\ket{\phi^{xy}}_{A'B'} &= \frac{1}{\sqrt{2}}(\ket{0}_a\ket{x\cdot y}_b + \ket{1}_a \ket{x \cdot y \oplus 1}_b)\\
	& = \frac{1}{\sqrt2}(\ket{\hat{z}}_a\ket{(-1)^{x.y}\hat{z}}_b + \ket{-\hat{z}}_a\ket{-(-1)^{x.y}\hat{z}}_b),
\end{align*}
where we have reordered the states in the tensor products to put Alice's states first. Moreover, we have mapped each qubits to vectors in corresponding Bloch spheres following:

\begin{equation*}
	\ket{\psi} = \cos \frac{\theta}{2} \ket{0} + e^{i\phi} \sin \frac{\theta}{2} \ket{1} \equiv \ket{\hat{n}} \mapsto (\theta,\phi) \mapsto \hat{n} \in \mathbb{S}^2
\end{equation*}
Now, 
\begin{align*}
	P(a,b|x,y) &= \bra{\phi^{xy}}\Pi^a_x\otimes \Pi^b_y\ket{\phi^{xy}}\\
	&= \frac{1}{\sqrt2}(\bra{\hat{z}}_a\bra{(-1)^{x.y}\hat{z}}_b + \bra{-\hat{z}}_a\bra{-(-1)^{x.y}\hat{z}}_b)(\frac{1}{2}\left \{\mathbb{I}+ a\hat{a}_x\cdot\hat{\sigma} \right\} \otimes\\& \quad \quad\quad \quad \quad \quad \frac{1}{2}\left \{\mathbb{I}+ b\hat{b}_y\cdot\hat{\sigma} \right\})\frac{1}{\sqrt2}(\ket{\hat{z}}_a\ket{(-1)^{x.y}\hat{z}}_b + \ket{-\hat{z}}_a\ket{-(-1)^{x.y}\hat{z}}_b)\\
	&= \frac{1}{8}\left\{ (1 + a\bra{\hat{z}}\hat{a}_x\cdot\hat\sigma\ket{\hat{z}}_a) \otimes (1 + b\;\bra{(-1)^{x\cdot y}\hat{z}}\hat{b}_y\cdot\hat\sigma\ket{(-1)^{x\cdot y}\hat{z}}_b)\right.\\
	&\quad \quad \quad \left.+ (a\bra{\hat{z}}\hat{a}_x\cdot\hat\sigma\ket{-\hat{z}}_a) \otimes (b\;\bra{(-1)^{x\cdot y}\hat{z}}\hat{b}_y\cdot\hat\sigma\ket{-(-1)^{x\cdot y}\hat{z}}_b) \right.\\
	&\quad \quad \quad \left.+ (a\bra{-\hat{z}}\hat{a}_x\cdot\hat\sigma\ket{\hat{z}}_a) \otimes (b\;\bra{-(-1)^{x\cdot y}\hat{z}}\hat{b}_y\cdot\hat\sigma\ket{(-1)^{x\cdot y}\hat{z}}_b) \right.\\
	&+ \left. (1 + a\bra{-\hat{z}}\hat{a}_x\cdot\hat\sigma\ket{-\hat{z}}_a) \otimes (1 + b\;\bra{-(-1)^{x\cdot y}\hat{z}}\hat{b}_y\cdot\hat\sigma\ket{-(-1)^{x\cdot y}\hat{z}}_b)\right.\}
\end{align*}
In the last expression, we note that we have two kind of terms --- 
\begin{enumerate}
	\item $a\bra{s\hat{z}}\hat{a}_x\cdot\hat\sigma\ket{s\hat{z}}_a$, with $s\in\{-1,1\}$ which compactly represents $a\bra{0}\hat{a}_x\cdot\hat\sigma\ket{0}_a$ when $s=1$, and $a\bra{1}\hat{a}_x\cdot\hat\sigma\ket{1}_a$ when $s=-1$.
	\item Next kind of term is $a\bra{s\hat{z}}\hat{a}_x\cdot\hat\sigma\ket{-s\hat{z}}_a$ representing $a\bra{0}\hat{a}_x\cdot\hat\sigma\ket{1}_a$ or $a\bra{1}\hat{a}_x\cdot\hat\sigma\ket{0}_a$.
\end{enumerate}
Now,
\begin{align}
	a\bra{0}\hat{a}_x\cdot\hat\sigma\ket{0}_a &=a \bra{0}a^x_1\sigma_x + a^x_2\sigma_y + a^x_3\sigma_z\ket{0}= a\;a^x_3,\nonumber\\
	a\bra{1}\hat{a}_x\cdot\hat\sigma\ket{1}_a &=a \bra{1}a^x_1\sigma_x + a^x_2\sigma_y + a^x_3\sigma_z\ket{1}= -a\;a^x_3;\nonumber\\
	\Rightarrow~
	&\mathbf{a\bra{s\hat{z}}\hat{a}_x\cdot}\hat\sigma \mathbf{\ket{s\hat{z}}_a} \mathbf{=s a\;a^x_3}. \label{eq: first-term}
\end{align}

Similarly, 
\begin{align}
	a\bra{0}\hat{a}_x\cdot\hat\sigma\ket{1}_a &=a \bra{0}a^x_1\sigma_x + a^x_2\sigma_y + a^x_3\sigma_z\ket{1}= a\;(a^x_1 -ia^x_2),\nonumber\\
	a\bra{1}\hat{a}_x\cdot\hat\sigma\ket{0}_a &=a \bra{1}a^x_1\sigma_x + a^x_2\sigma_y + a^x_3\sigma_z\ket{0}= a\;(a^x_1 +ia^x_2);\nonumber\\
	&\Rightarrow~\mathbf{a\bra{s\hat{z}}\hat{a}_x\cdot\ }\hat{\sigma}\mathbf{\ket{-s\hat{z}}_a}=\mathbf{a\;(a^x_1 -s\;ia^x_2)}.\label{eq: second-term}
\end{align}
Using Eqs.~(\ref{eq: first-term}), and (\ref{eq: second-term}), we obtain
\begin{align}
	P(a,b|x,y) &= \frac{1}{8}\left\{ (1 + a\;a^x_3) \otimes (1 + (-1)^{xy} b \;b^y_3)+ (a\;(a^x_1 -ia^x_2)) \otimes (b\;(b^y_1 -(-1)^{xy}\;ib^y_2)) \right.\nonumber\\
	&\quad \quad \quad \left.+ (a\;(a^x_1 +ia^x_2)) \otimes (b\;(b^y_1 +(-1)^{xy}\;ib^y_2)) + (1 - a\;a^x_3) \otimes (1 - (-1)^{xy} b \;b^y_3)\right.\}\nonumber\\
	&= \frac{1}{8}\left\{2(1 + ab(-1)^{xy}\;a^x_3 b^y_3) + 2ab \; (a^x_1 b^y_1 - (-1)^{xy}\;a^x_2 b^y_2) \right \}\nonumber\\
	&= \mathbf{\frac{1}{4}\left \{ 1 + ab\; (a^x_1b^y_1 - (-1)^{xy}\;a^x_2 b^y_2+ (-1)^{xy}\;a^x_3 b^y_3)\right\}}.\label{eq: final_prob}
\end{align}
\noindent Clearly, Eq.~(\ref{eq: final_prob}) satisfies
\begin{equation}
	P(a \mid x,y) = \sum _b P(a,b\mid x,y) = \frac{1}{2} = P(a \mid x,y').
\end{equation}
Similarly, $P(b \mid x,y) =  \frac{1}{2} = P(b \mid x',y)$. Hence, \textbf{the quantum state $\ket{\phi^{xy}}$ is indeed a no-signaling resource.}

We now focus our attention to the unentangled state $\rho_c^\star$, and find out if it's no-signaling. For this, we look at the corresponding mixed density matrix $\rho^\star_c$. Precisely, 

\begin{align}
	\rho^\star_c &= \frac{1}{2}\left \{ \proj{0}_A \otimes \proj{x\cdot y}_B + \proj{1}_A \otimes \proj{1 \oplus x\cdot y}_B \right \}. 
\end{align}

The probability $P(a,b \mid x,y)$ is then given by
\begin{align}
	P(a,b \mid x,y) &= \Tr[ \Pi^x_a \otimes \Pi^y_b \; \rho^\star_c]\nonumber\\
	&= \frac{1}{8} \Tr \left [ \{(\mathbb{I}+ a \;\hat{a}_x \cdot \hat{\sigma})\otimes (\mathbb{I} + b \; \hat{b}_y \cdot \hat{\sigma})\} \{\proj{0}_A \otimes \proj{x\cdot y}_B + \proj{1}_A \otimes \proj{1 \oplus x\cdot y}_B \} \right ]\nonumber\\
	&= \frac{1}{8} \Tr \left [ (\mathbb{I}+ a \;\hat{a}_x \cdot \hat{\sigma}) \proj{0}_A \right ] \Tr \left [(\mathbb{I} + b \; \hat{b}_y \cdot \hat{\sigma}) \proj{x\cdot y}_B \right ]\nonumber\\
	& \quad \quad + \frac{1}{8} \Tr \left [ (\mathbb{I}+ a \;\hat{a}_x \cdot \hat{\sigma}) \proj{1}_A \right ] \Tr \left [(\mathbb{I} + b \; \hat{b}_y \cdot \hat{\sigma}) \proj{1 \oplus x\cdot y}_B \right ].
	\label{eq: semifinal_prob_classical}
\end{align}
Upon simplifying the last expression, we get terms like: $\Tr[(a\; \hat{a}_x\cdot \hat{\sigma}) \proj{0}_A]$ and $\Tr[(a\; \hat{a}_x\cdot \hat{\sigma}) \proj{1}_A]$ which can be condensed into the form $\Tr[(a\; \hat{a}_x\cdot \hat{\sigma}) \proj{s \hat{z}}_A]$, where $s = 1 \Rightarrow \proj{0}$ and $s = -1 \Rightarrow \proj{1}$. We have 
\begin{equation}
	\Tr[(a\; \hat{a}_x\cdot \hat{\sigma}) \proj{s \hat{z}}_A] = sa\; a^x_3.
	\label{eq: condensed_trace}
\end{equation}
Using Eq.~(\ref{eq: condensed_trace}) in Eq.~(\ref{eq: semifinal_prob_classical}) and taking the traces, we get
\begin{align}
	P(a,b\mid x,y) &= \frac{1}{8} \left ( 1 + a \; a^x_3\right )\left ( 1 + (-1)^{x\cdot y} b \; b^y_3\right ) + \frac{1}{8} \left ( 1 - a \; a^x_3\right )\left ( 1 - (-1)^{x\cdot y} b \; b^y_3\right )\nonumber\\
	&= \mathbf{\frac{1}{4}\left [ 1 + (-1)^{x\cdot y}\; ab \; a^x_3b^y_3\right]}.
	\label{eq: final_classical_prob}
\end{align}
Again, Eq.~(\ref{eq: final_classical_prob}) holds the relation
\begin{equation}
	P(a \mid x,y) = \sum _b P(a,b\mid x,y) = \frac{1}{2} = P(a \mid x,y').
\end{equation}
Similarly, $P(b \mid x,y) =  \frac{1}{2} = P(b \mid x',y)$. Thus, \textbf{the classical state $\rho_c^{\star} = \rho^{xy}$ too is a no-signaling resource.} Moreover, both quantum and classical expressions for $P(a,b \mid x,y)$ cannot be factorized as 
\begin{equation}
	P(a, b\mid x,y) \ne P(a\mid x) P(b \mid y),
\end{equation}
hence both $\ket{\phi^{xy}}$ and $\rho^{xy}$ are \textbf{nonlocal resources.} We now turn our attention to finding the appropriate basis in which these states gives PR behavior, i.e., when $\mathcal{P}^\star = \mathcal{P}_{PR}$.

\section{B. Allowed measurement bases to obtain PR correlations using $\ket{\phi^{xy}}$ and $\rho^{xy}$}

In the Letter, we said that PR behavior is characterized by the following probability equation:
\begin{equation}
	\mathcal{P}_{PR}: \; P(a,b\mid x,y)
	= 
	\begin{cases}
		\displaystyle \tfrac12, & \text{when} \quad a\cdot b = (-1)^{x\,y}\\[6pt]
		0,                      & \text{otherwise}.
	\end{cases}
	\label{eq: PR-conditions}
\end{equation}

To get the measurement bases which yield PR correlations, which from now we call PR bases, we find directions $\hat{a}_x = (a^x_1, a^x_2, a^x_3)$ and $\hat{b}_y = (b^y_1, b^y_2, b^y_3)$ which satisfy Eq.~(\ref{eq: PR-conditions}). For the classical case, putting $a\cdot b = (-1)^{x\cdot y}$ in Eq.~(\ref{eq: final_classical_prob}) and equating it to 0.5, we get:
\begin{align}
	P(a,b \mid x,y) &= \frac{1}{4}\left [ 1 +  a^x_3b^y_3\right] = \frac{1}{2}~\Rightarrow a^x_3b^y_3= 1 \nonumber
	\\
	&\Rightarrow (a^x_3, b^y_3) \in \{(1,1); (-1,-1)\}.
\end{align}
Similarly, for the case when $a\cdot b \ne (-1)^{x\cdot y}$, we equate
\begin{align}
	P(a,b \mid x,y) &= \frac{1}{4}\left [ 1 -  a^x_3b^y_3\right] =0~\Rightarrow a^x_3b^y_3= 1 \nonumber\\
	&\Rightarrow(a^x_3, b^y_3) \in \{(1,1); (-1,-1)\}.
\end{align}

Hence, in both the cases of Eq.~(\ref{eq: PR-conditions}), for the classical state $\rho^{\star}_c$, we get solutions $\hat{a}_x = (0,0,\pm1)$ and $\hat{b}_y = (0, 0, \pm1)$ which maps to the \textbf{computational bases} $\{\ket{0}, \ket{1} \}$. Consequently,
\begin{equation}
	\boxed{
		\text{For } \rho^{xy}, \mathcal{P^\star} = \mathcal{P}_{PR} \text{ only in basis } \{\ket{0}, \ket{1}\}.}
\end{equation}

We now find the PR bases for the quantum state $\rho^\star_q$. Putting conditions in  Eq.~(\ref{eq: PR-conditions}) to the quantum expression for $P(a,b \mid x,y)$ [Eq.~(\ref{eq: final_prob})], we have the following two cases: (i) $ab = (-1)^{xy}$, and (ii) $ab \ne (-1)^{xy}$.
\begin{table}[h!]
	\centering
	\begin{tabular}{|c|c|c|c|}
		\hline
		~~~$\mathbf{x}$~~~ & ~~~$\mathbf{y}$~~~ & $\mathbf{ab}$ ($ab = (-1)^{xy}$) & $\mathbf{ab}$ ($ab \ne (-1)^{xy}$) \\
		\hline
		0 & 0 & +1 & -1 \\
		\hline
		0 & 1 & +1 & -1 \\
		\hline
		1 & 0 & +1 & -1 \\
		\hline
		1 & 1 & -1 & +1 \\
		\hline
	\end{tabular}
	\caption{PR correlations}
	\label{tab: PR}
\end{table}

Recall that for inputs $(0,0),(0,1)$, and $(1,0)$, the corresponding projector sets $(\mathcal{M}^0_A,\mathcal{M}^0_B), (\mathcal{M}^0_A,\mathcal{M}^1_B)$, and $(\mathcal{M}^1_A,\mathcal{M}^0_B)$ are mapped to vectors $(\hat{a}_0,\hat{b}_0),(\hat{a}_0,\hat{b}_1)$, and $(\hat{a}_1,\hat{b}_0)$, respectively. Also for the same inputs, when $ab = (-1)^{xy}$, we have product of outputs $ab = 1$ (Table~\ref{tab: PR}). Equation~(\ref{eq: final_prob}), and (\ref{eq: PR-conditions}) together give
\begin{align}
	P(a,b|x',y') = \frac{1}{4}\left \{ 1 +  (a^{x'}_1b^{y'}_1 - a^{x'}_2 b^{y'}_2+ a^{x'}_3 b^{y'}_3)\right\}&=\frac{1}{2}\nonumber\\
	\Rightarrow \mathbf{a^{x'}_1b^{y'}_1 - a^{x'}_2 b^{y'}_2+ a^{x'}_3 b^{y'}_3 - 1} &=\mathbf{0}, \label{eq: 1}
\end{align}
where $(x', y') \in \{(0,0),(0,1), (1,0)\}$. For input $(1,1)$, $(\mathcal{M}^1_A,\mathcal{M}^1_B) \mapsto (\hat{a}_1,\hat{b}_1)$, and when $ab = (-1)^{xy}$ we have $ab = -1$. In this case, Eqs.~(\ref{eq: final_prob}) and (\ref{eq: PR-conditions}) give
\begin{align}
	P(a,b|1,1) = \frac{1}{4}\left \{ 1 -  (a^{1}_1b^{1}_1 + a^{1}_2 b^{1}_2- a^{1}_3 b^{1}_3)\right\}&=\frac{1}{2}\nonumber\\
	\Rightarrow \mathbf{a^{1}_1b^{1}_1 + a^{1}_2 b^{1}_2- a^{1}_3 b^{1}_3 + 1} &=\mathbf{0}. \label{eq: 2}
\end{align}

When we put the other condition i.e $ab \ne (-1)^{xy} \Rightarrow P(a,b|x,y) = 0$ and solve for bases corresponding to every input combination (Table \ref{tab: PR}), we again get Eq.~(\ref{eq: 1}) for inputs $(0,0),(0,1)$, and $(1,0)$ and Eq.~(\ref{eq: 2}) for input $(1,1)$, respectively.

We are thus left with to solve Eqs.~(\ref{eq: 1}), and (\ref{eq: 2}). Recall that 
\begin{align}
	\hat{a}_x &= (a^x_1,a^x_2,a^x_3) = (\;\sin(\theta^x_a) \cos(\phi^x_a),\; \sin(\theta^x_a) \sin(\phi^x_a),\; \cos(\theta^x_a)\;), \label{eq: alice_basis}\\ \hat{b}_y &= (b^y_1,b^y_2,b^y_3) = (\;\sin(\theta^y_b) \cos(\phi^y_b),\; \sin(\theta^y_b) \sin(\phi^y_b),\; \cos(\theta^y_b)\;).\label{eq: bob_basis}
\end{align}
Substituting Eq.~(\ref{eq: alice_basis}) in Eq.~(\ref{eq: 1}) and Eq.~(\ref{eq: bob_basis}) in Eq.~(\ref{eq: 2}) we obtain,
\begin{align}
	& \quad \quad \quad \quad \quad \quad \quad \quad \quad \quad \quad \quad \quad \; \mathbf{a^{x'}_1b^{y'}_1 - a^{x'}_2 b^{y'}_2+ a^{x'}_3 b^{y'}_3 - 1} =\mathbf{0}\nonumber\\
	&\Rightarrow\sin(\theta^{x'}_a) \cos(\phi^{x'}_a)\sin(\theta^{y'}_b) \cos(\phi^{y'}_b)- \sin(\theta^{x'}_a) \sin(\phi^{x'}_a)\sin(\theta^{y'}_b) \sin(\phi^{y'}_b) + \cos(\theta^{x'}_a) \cos(\theta^{y'}_b) = 1\nonumber\\
	&\Rightarrow\sin(\theta^{x'}_a)\sin(\theta^{y'}_b)\{\cos(\phi^{x'}_b + \phi^{y'}_a)\} + \cos(\theta^{x'}_a) \cos(\theta^{y'}_b) = 1,\label{eq: first_trig_eq}\\
	& \quad \quad \quad \quad \quad \quad \quad \quad \quad \quad \quad \quad \quad \; \mathbf{a^{1}_1b^{1}_1 + a^{1}_2 b^{1}_2- a^{1}_3 b^{1}_3 + 1} =\mathbf{0}\nonumber\\
	&\Rightarrow\sin(\theta^{1}_a) \cos(\phi^{1}_a)\sin(\theta^{1}_b) \cos(\phi^{1}_b)+ \sin(\theta^{1}_a) \sin(\phi^{1}_a)\sin(\theta^{1}_b) \sin(\phi^{1}_b) - \cos(\theta^{1}_a) \cos(\theta^{1}_b) = - 1\nonumber\\    
	&\Rightarrow\sin(\theta^{1}_a)\sin(\theta^{1}_b)\{\cos(\phi^{1}_b - \phi^{1}_a)\} - \cos(\theta^{1}_a) \cos(\theta^{1}_b) = -1,\label{eq: second_trig_eq}
\end{align}
where $(x', y') \in \{(0,0),(0,1), (1,0)\}$. The solutions of Eqs.~(\ref{eq: first_trig_eq}) and (\ref{eq: second_trig_eq}) are listed in Table \ref{tab: first_trig_eq} and Table \ref{tab: second_trig_eq}, respectively. Computational basis of course is a common solution. The novel bases must be an intersection of the two solution sets given by second row of Tables~\ref{tab: first_trig_eq} and \ref{tab: second_trig_eq}. Thus, $2\pi - l_1=\phi^1_b = \pi + l_2$, or $l_1 + l_2 = \pi$. But $l_1 = \phi^1_a = l_2$; hence, $l_1 = l_2 = \pi/2$.\\
\begin{table}[h!]
	\centering
	\begin{tabular}{|c|c|c|c|}
		\hline
		$\theta^0_a = \theta^1_a = \theta^0_b = \theta^1_b = \{0,\pi \}$ & $\phi^0_a, \phi^1_a \in [0, 2\pi]$ & $\phi^0_b , \phi^1_b \in [0, 2\pi]$ & Computational Basis \\
		\hline
		$\theta^0_a = \theta^1_a = \theta^0_b = \theta^1_b = k \in [0, \pi]$ & $\phi^0_a= \phi^1_a = l_1\in [0, 2\pi]$ & $\phi^0_b = \phi^1_b = 2\pi -l_1$ & Novel Bases \\
		\hline
	\end{tabular}
	\caption{Solutions of Eq.~(\ref{eq: first_trig_eq})}
	\label{tab: first_trig_eq}
\end{table}

\begin{table}[h!]
	\centering
	\begin{tabular}{|c|c|c|c|}
		\hline
		$\theta^1_a = \theta^1_b = \{0,\pi\}$ & $ \phi^1_a \in [0, 2\pi]$ & $\phi^1_b \in [0, 2\pi]$ & Computational Basis \\
		\hline
		$\theta^1_a =\theta^1_b = k \in [0, \pi]$ & $\phi^1_a = l_2\in [0, 2\pi]$ & $\phi^1_b = \pi +l_2$ & Novel Bases \\
		\hline
	\end{tabular}
	\caption{Solutions of Eq.~(\ref{eq: second_trig_eq})}
	\label{tab: second_trig_eq}
\end{table}

\begin{figure}[h!]
	\centering
	\includegraphics[width=0.4\linewidth]{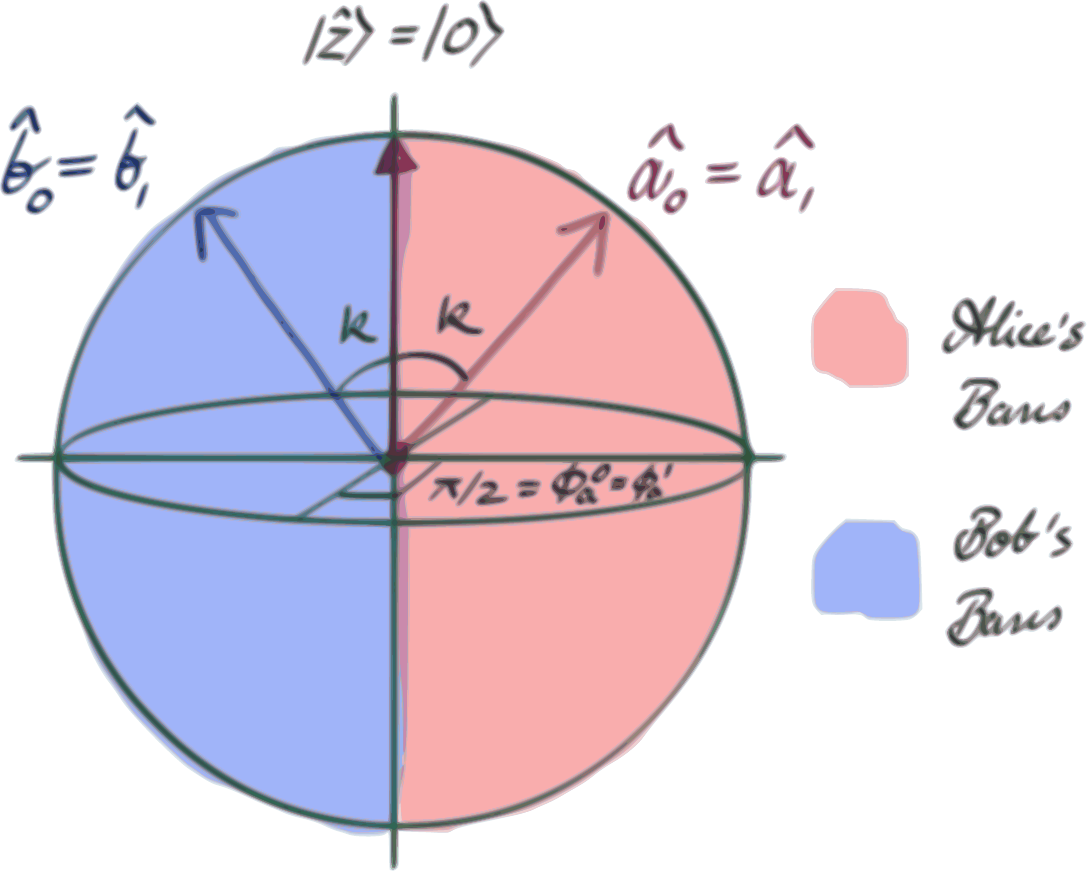}
	\caption{PR bases for quantum state $\ket{\phi^{xy}}$ on Bloch sphere.}
	\label{fig: sol.Bloch}
\end{figure}

\textbf{Result}: The common solution, alternatively, the PR bases for the quantum state $\ket{\phi^{xy}}$ is shown in Table~\ref{tab: common_sol}. Figure~\ref{fig: sol.Bloch} shows the bases satisfying PR correlations on the Bloch sphere.

\begin{table}[h!]
	\centering
	\begin{tabular}{|c|c|c|c|}
		\hline
		$\theta^0_a = \theta^1_a = \theta^0_b = \theta^1_b = \{0, \pi \}$ & $\phi^0_a, \phi^1_a \in [0, 2\pi]$ & $\phi^0_b , \phi^1_b \in [0, 2\pi]$ & Computational Basis \\
		\hline
		$\theta^0_a = \theta^1_a = \theta^0_b = \theta^1_b = k \in [0, \pi]$ & $\phi^0_a= \phi^1_a = \pi/2$ & $\phi^0_b = \phi^1_b = 3\pi/2$ & Novel Basis \\
		\hline
	\end{tabular}
	\caption{PR bases for quantum state $\ket{\phi^{xy}}$.}
	\label{tab: common_sol}
\end{table}

Alternatively, 
\begin{equation}
	\boxed{
		\text{For } \ket{\phi^{xy}}, \mathcal{P^\star} = \mathcal{P}_{PR} \text{ only in measurement bases given by Table~\ref{tab: common_sol}.}}
\end{equation}

\section{C. Theory supporting results in Fig.~3 of the main manuscript}
Figure 3(a) of the main text shows experimentally obtained CHSH values when Alice and Bob perform local measurements in computational, diagonal, and circular bases. The bar graph shows that computational and circular bases realise PR correlations, whereas diagonal bases do not. Having derived PR bases in the previous section, we can see that the computational basis  $\{\ket{0}, \ket{1}\}\mapsto\{\ket{H}, \ket{V}\}$ clearly lies in the set of PR bases (Fig.~\ref{fig: sol.Bloch}, Tab.~\ref{tab: common_sol}). Next, we see that the circular basis actually refers to measurement bases $\hat{a}_x \mapsto (\theta^x_a, \phi^x_a) = (\pi/2, \pi/2)$, for both input values, i.e. $x \in \{0,1\}$. However, Bob uses the same bases but assigns opposite output values, namely, his measurement bases are $\hat{b}_y \mapsto (\theta^y_b, \phi^y_b) = (\pi/2, 3\pi/2)$, where $y\in \{0,1\}$. Hence, the circular basis with the given associations belongs to the \emph{
	novel basis} showed in Table~\ref{tab: common_sol}. The diagonal basis does not belong to the PR bases set, and hence, as expected, it cannot yield PR correlations. For clarity, we restate the measurement protocols used by Alice and Bob for the three bases: 
\begin{enumerate}
	\item Both Alice and Bob measure \(\ket{\phi^{xy}}_{B'A'}\) in the computational basis, assigning \(a=-1\) (\(b=-1\)) when outcome \(\ket{0}\mapsto\ket{H}\) is observed and \(a=+1\) (\(b=+1\)) otherwise.
	\item Both measure in the \(\{\ket{+},\ket{-}\}\mapsto\{\ket{A},\ket{D}\}\) basis, assigning \(a=-1\) (\(b=-1\)) when \(\ket{+}\mapsto\ket{D}\) is observed and \(a=+1\) (\(b=+1\)) otherwise.
	\item Both measure in the circular basis \(\{\ket{L},\ket{R}\}\), with Alice assigning \(a=-1\) for \(\ket{L}\) and \(a=+1\) for \(\ket{R}\), while Bob assigns the opposite outcomes \(b=+1\) for \(\ket{L}\) and \(b=-1\) for \(\ket{R}\).
\end{enumerate}

We have Eq.~(\ref{eq: final_prob}) which gives the correlation between inputs $a,b$ and outputs $x,y$ for any arbitrary choice of measurement bases. In the following, we derive the CHSH score $\mathcal{S}$ for any arbitrary measurement basis using Eq.~(\ref{eq: final_prob}). We have $\mathcal{S} = \braket{a_0b_0}+ \braket{a_0b_1}+\braket{a_1b_0}-\braket{a_1b_1}$

Now, 
\begin{align*}
	\braket{a_xb_y} &= \sum_{a,b} ab \; P(a,b \mid x,y)= P(1,1\mid x,y) - P(1,-1\mid x,y) - P(-1,1\mid x,y) + P(-1,-1\mid x,y)\\
	&=  (a^x_1b^y_1 - (-1)^{xy}\;a^x_2 b^y_2+ (-1)^{xy}\;a^x_3 b^y_3),
\end{align*}
giving,
\begin{align}
	\mathcal{S} &= \braket{a_0b_0}+ \braket{a_0b_1}+\braket{a_1b_0}-\braket{a_1b_1}\nonumber\\
	&= (a^0_1b^0_1 - a^0_2 b^0_2+ a^0_3 b^0_3) + (a^0_1b^1_1 - a^0_2 b^1_2+ a^0_3 b^1_3) + (a^1_1b^0_1 - a^1_2 b^0_2+ a^1_3 b^0_3) - (a^1_1b^1_1 + a^1_2 b^1_2- a^1_3 b^1_3)\nonumber\\
	&= a^0_1(b^0_1 + b^1_1) - a^0_2(b^0_2 + b^1_2) + a^0_3(b^0_3 + b^1_3)+ a^1_1(b^0_1 - b^1_1) - a^1_2(b^0_2 + b^1_2) + a^1_3(b^0_3 + b^1_3)\nonumber\\
	&=\sin\theta^0_a \cos\phi^0_a(\sin\theta^0_b \cos\phi^0_b + \sin\theta^1_b \cos\phi^1_b) - \sin\theta^0_a \sin\phi^0_a(\sin\theta^0_b \sin\phi^0_b + \sin\theta^1_b \sin\phi^1_b)\nonumber\\ 
	&\quad\quad+ \cos\theta^0_a(\cos\theta^0_b + \cos\theta^1_b)+\sin\theta^1_a \cos\phi^1_a(\sin\theta^0_b \cos\phi^0_b - \sin\theta^1_b \cos\phi^1_b)\nonumber\\
	&\quad\quad - \sin\theta^1_a \sin\phi^1_a(\sin\theta^0_b \sin\phi^0_b + \sin\theta^1_b \sin\phi^1_b) + \cos\theta^1_a(\cos\theta^0_b + \cos\theta^1_b).
	\label{eq: final_S}
\end{align}

\noindent The last equality is obtained by the associations given in Eq.~(\ref{eq: alice_basis}) and Eq.~(\ref{eq: bob_basis}). Putting $\theta^0_a = \theta^1_a = \theta^0_b = \theta^1_b = \{0, \pi \}$ in Eq.~(\ref{eq: final_S}), we get $\mathcal{S} = 4$ in the computational basis as expected. Moreover, putting $\theta^0_a = \theta^1_a = \theta^0_b = \theta^1_b = \pi/2$, and $\phi^0_a= \phi^1_a = \pi/2$, and $\phi^0_b = \phi^1_b = 3\pi/2$, we get $\mathcal{S} = 4$, further confirming PR correlations in the circular measurement basis. Next, in the diagonal basis, i.e, when $\theta^0_a = \theta^1_a = \theta^0_b = \theta^1_b = \pi/2$, and $\phi^0_a= \phi^1_a = \phi^0_b = \phi^1_b = 0$, we get $\mathcal{S} = 2$, which is what we got experimentally as well (Fig.~3(a), main text). 

Lastly, for the entangled state $\ket{\phi^{xy}}$, when we put $\phi^0_a= \phi^1_a = \phi^0_b = \phi^1_b = 0$, and vary the polar angles, i.e., $\theta^0_a = \theta^1_a = \theta^0_b = \theta^1_b = \lambda$, we get $\mathcal{S}(\lambda) = 2 + 2 \cos^2\lambda$. In Fig.~3(b) of the main text, the theoretical plot is exactly the curve $\mathcal{S}(\lambda)$ supported by the experimental results.

For the mixed unentangled state $\rho^{xy}$, we can do the same analysis beginning from Eq.~(\ref{eq: final_classical_prob}) to get $\mathcal{S}$ for any arbitrary measurement bases. We get

\begin{equation}
	\mathcal{S} = (\cos\theta^0_a + \cos\theta^1_a)(\cos\theta^0_b+\cos\theta^1_b),
\end{equation}
which illustrates the fact that for $\rho^{xy}$, unlike $\ket{\phi^{xy}}$, $\mathcal{S}$ is independent of the azimuthal angle $\phi$. This is also evident in the last bar of Fig.~3(c), where $\mathcal{S}$ for both diagonal and circular basis comes out to be the same. Putting $\theta^0_a = \theta^1_a = \theta^0_b = \theta^1_b = 0$, we get $\mathcal{S} = 4$; while putting $\theta^0_a = \theta^1_a = \theta^0_b = \theta^1_b = \pi/6$ and $\theta^0_a = \theta^1_a = \theta^0_b = \theta^1_b = \pi/4$ yields $\mathcal{S} = 3$, and 4, respectively, which are also supported by the numerical simulation given in Fig.~3(c). 

\section{D. Entanglement Source Preparation and Characterization} 
A bright entangled photon source is prepared employing the type-II Sponteneous Parametric Down Conversion (SPDC) process within the Sagnac geometry, that generates degenerate pairs of polarization-entangled photons at the wavelength $808.5$ nm. The pump beam is prepared in the polarization state $\ket{D} = \frac{1}{\sqrt{2}} \left( \lvert H \rangle + \lvert V \rangle \right)$, which is then focused onto the non-linear PPKTP crystal placed within the Sagnac interferometer using an achromatic lens ($L_1$) of focal length 20 cm. The interferometer is build using a dual-band polarizing beam splitter ($D-PBS$) and two mirrors ($M_1$, $M_2$). The pump beam when incident on the $D-PBS$, gets splitted into two paths based on polarization: The horizontally polarized ($\ket{H}$) component gets transmitted and propagates clockwise (CW) within the interferometer, while the vertically polarized ($\ket{V}$) component gets reflected and propagate counter-clockwise (CCW). A half-wave plate ($HWP_{2}$) with its fast axis aligned at $45^{\circ}$ w.r.to horizontal converts the $V$-polarization into $H$-polarization to satisfy the type-II quasi-phase-matching condition in the nonlinear PPKTP crystal. \\

The $H$-polarized pump photons from both the sides are made incident on the PPKTP crystal kept within a oven mounted on a 1D-translational stage (along z) between $M_{1}$ and $M_{2}$ of the interferometer. Within the crystal, H-polarized pump photons probabilistically down converts into a pair of daughter photons in $H$ and $V$ polarizations, respectively. The set at the temperature $30.9^{\circ}~C$, at which the degeneracy is obtained for the particular PPKTP crystal. The crystal is periodically poled with a \( 10~\mu\text{m} \) grating period to ensure efficient degenerate down-conversion (SPDC). The down-converted photons are collimated using the lenses $L_1$ and $L_2$ of focal length $20$ cm and collected into the single-mode fibers using the couplers. The polarization controllers compensate for any relative phase introduced due to fiber birefringence. At the output end of the two fibers, the resultant state becomes
\begin{align}
	\ket{\Psi^{+}} = \frac{1}{\sqrt{2}} \left( \ket{H}\ket{V} + \ket{V}\ket{H} \right).
\end{align}

The above state is a maximally entangled state in polarization degree of freedom. The visibilities are measured for the prepared state in different polarization basis. The measured visibilities are \(99.72\%\) and \(99.73\%\) for the idler in \(H\) and \(V\) bases, respectively, and \(98.77\%\) and \(98.83\%\) for the idler in \(D\) and \(A\) bases, respectively. The Bell violation value or the CHSH value obatined for the state is $S = 2.708\pm0.02$.  \\

\begin{figure}[h!]
	\centering
	\begin{minipage}{0.45\linewidth}
		\centering
		\includegraphics[width=\linewidth]{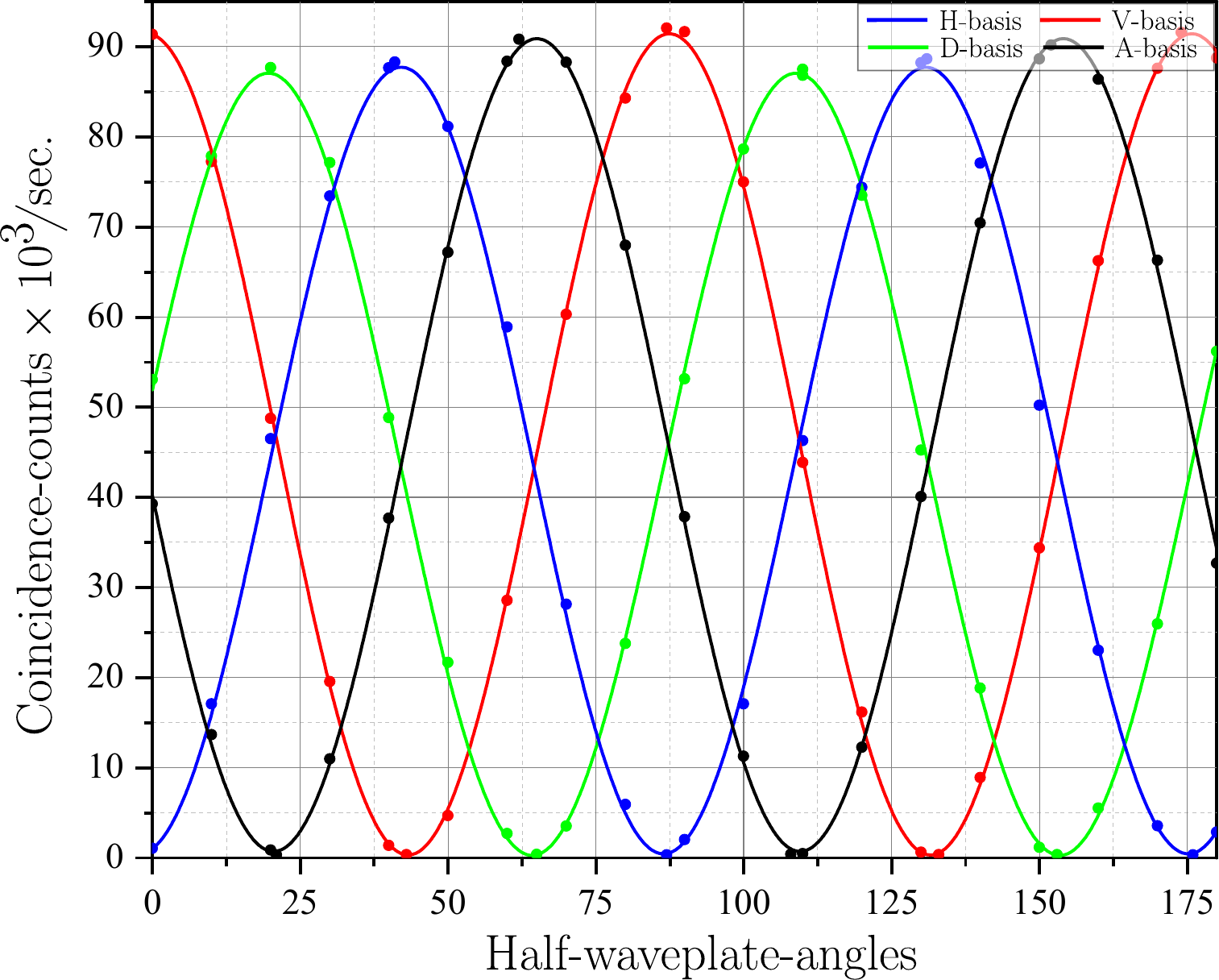}
		\label{fig:visibility}
	\end{minipage} \hfill
	\begin{minipage}{0.54\linewidth}
		\centering
		\includegraphics[width=\linewidth]{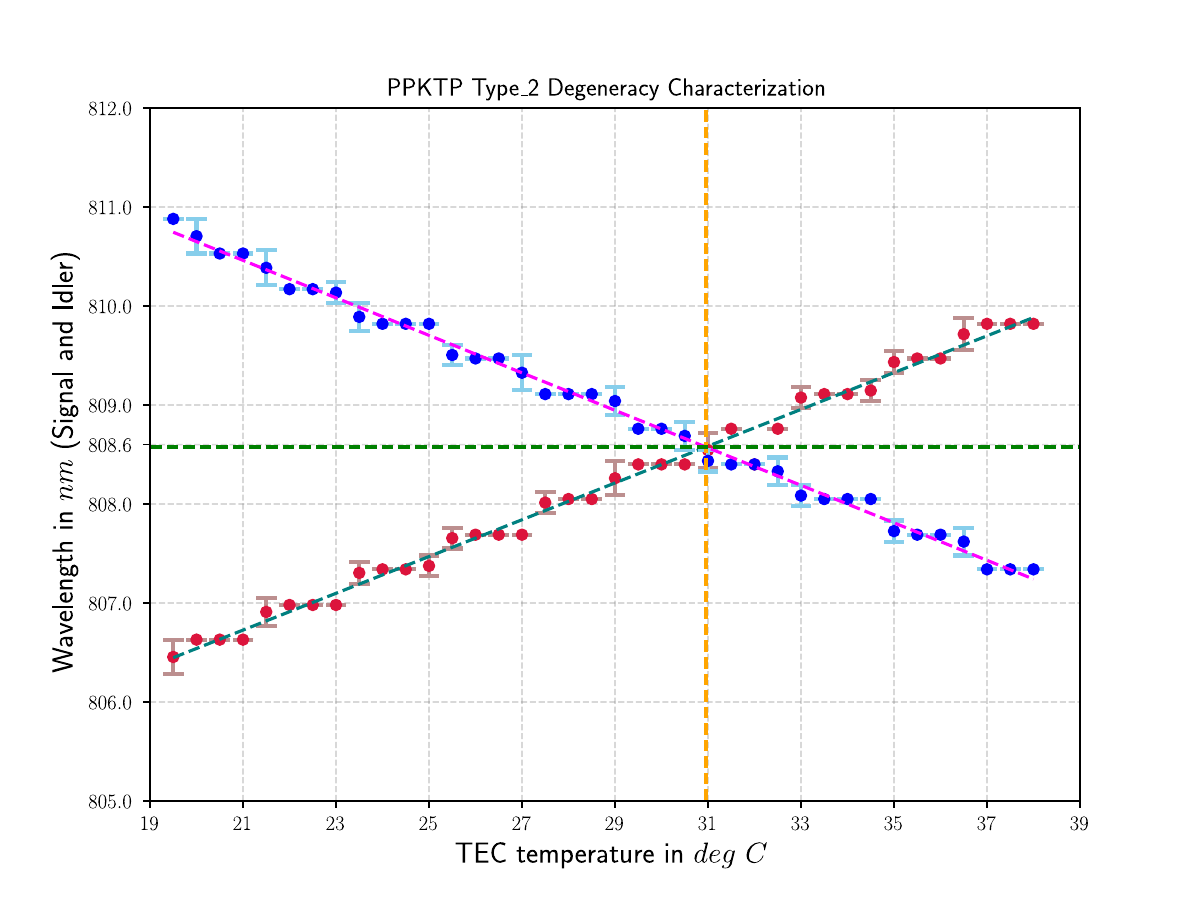}
		\label{fig:degeneracy}
	\end{minipage}
	\caption{Visibility and Degeneracy characterization of the source.}
	\label{fig: visibility}
\end{figure}

Further, the intensity correlation function \( g^{(2)}(\tau = 0) \) is determined to characterize the quality of the source. For this measurement, the idler photon is sent directly to the detector, labelled as reference channel ($R$) and the signal beam is split into two ($S_1$, $S_2$) using a 50:50 beam splitter. A three-fold coincidence measurement $C(S_{1}, S_{2} \mid R )$ is performed to evaluate \( g^{(2)}(0) \). For our prepared source, \( g^{(2)}(0) \) is obtained to be \( 0.004 \), which confirms the high purity and reliability of the single-photon source.\\

\section{E. Simulating multi-partite beyond quantum nonlocal correlations}
The study of Bell nonlocality naturally extends to the multipartite setting, where more than two distant parties—Alice, Bob, Charlie, $\cdots$—each receive inputs $x \in \mathcal{X},~ y \in \mathcal{Y},~ z \in \mathcal{Z}, \cdots$ and produce outputs $a \in \mathcal{A},~ b \in \mathcal{B},~ c \in \mathcal{C}, \cdots$. A multipartite input–output correlation $P \equiv {p(a,b,c,\cdots|x,y,z,\cdots)}$ is said to be no-signaling (NS) if no non-empty subgroup of parties can influence the marginal outcome probabilities of the remaining parties by varying their local inputs \hyperlink{1}{\textcolor{blue}{[1]}}. For finite input–output alphabets, the set $\mathcal{N}$ of all NS correlations forms a convex polytope embedded in a finite-dimensional Euclidean space. The polytope $\mathcal{N}$ has finitely many extremal points, including local deterministic ones as well as nonlocal indeterministic ones. Notably, none of the nonlocal extreme points of $\mathcal{N}$ is quantum realizable \hyperlink{3}{\textcolor{blue}{[3]}}.  

In the simplest bipartite $2-2-2$ scenario, the Popescu–Rohrlich (PR) correlation, up to local relabeling, is the only nonlocal extremal point. For scenarios involving more than two parties, however, qualitatively new types of genuinely multipartite extremal nonlocal correlations emerge. In what follows, we show that our simulation framework can be extended to such multipartite settings. Specifically, we analyze the $3-2-2$ Bell scenario, where three parties each perform two dichotomic measurements \hyperlink{1}{\textcolor{blue}{[1,}} \hyperlink{2}{\textcolor{blue}{2,}} \hyperlink{5}{\textcolor{blue}{5]}}. Even in this restricted case, different classes of genuine extremal correlations exist. Here, we focus on the subclass of correlations, known as the full correlation boxes whose one-party and two-party marginal distributions are uniform. Within this subclass, three distinct types of correlations are possible:
\begin{subequations}
	\begin{align}
		\text{XYZ-box:}&~~p(a,b,c|x,y,z)=\begin{cases}
			\tfrac{1}{4},~~~a\oplus b\oplus c=xyz;\\
			0,~~~\text{otherwise}.
		\end{cases}\label{xyz}\\
		\text{X(Y+Z)-box:}&~~p(a,b,c|x,y,z)=\begin{cases}
			\tfrac{1}{4},~~~a\oplus b\oplus c=xy\oplus  xz;\\
			0,~~~\text{otherwise}.
		\end{cases}\label{x(y+z)}\\
		\text{Svetlichny-box:}&~~p(a,b,c|x,y,z)=\begin{cases}
			\tfrac{1}{4},~~~a\oplus b\oplus c=xy\oplus yz\oplus zx;\\
			0,~~~\text{otherwise}.
		\end{cases}\label{Svetlichny}
	\end{align}   
\end{subequations}

\begin{figure}[H]
	\centering
	\includegraphics[width=.95\linewidth]{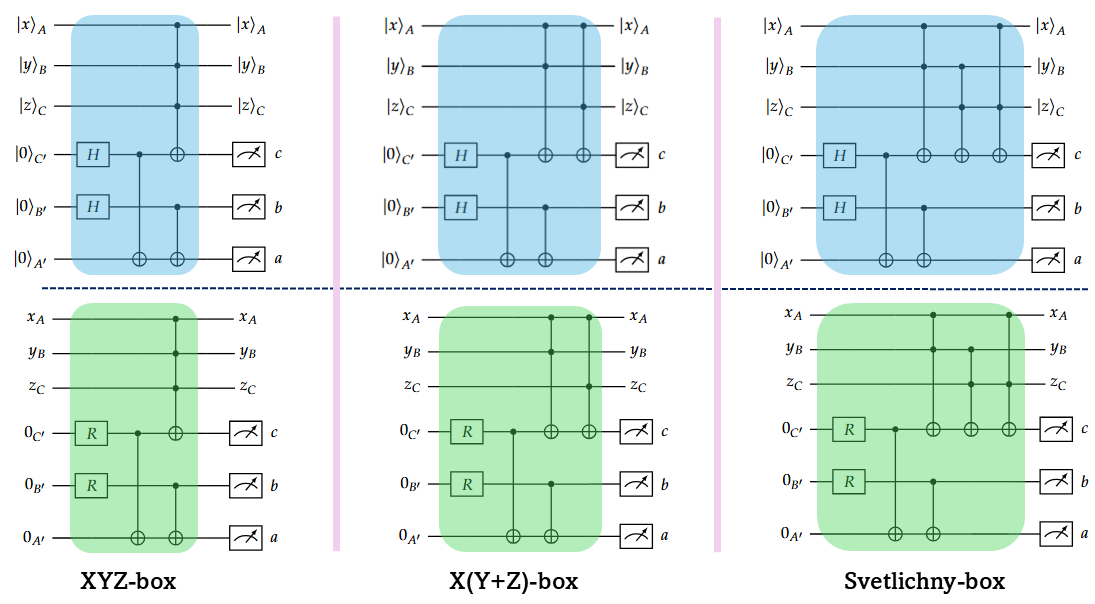}
	\caption{Top: Quantum oracles (blue shaded box). The three oracles from left to right simulate the XYZ-box, X(Y+Z)-box, and Svetlichny-box respectively. All the oracles satisfy NS conditions when the inputs for primed systems are fixed to $\ket{0}$ and the inputs for unprimed systems are limited to computational states only. Bottom: The corresponding classical oracles (green shaded box). }
	\label{figS1}
\end{figure}

\noindent {\bf Simulating the correlations with quantum/classical NS oracle:} The XYZ-correlation of Eq.(\ref{xyz}) extends the bipartite PR correlation to the tripartite setting. While PR correlations trivialize bipartite communication complexity problems \hyperlink{4}{\textcolor{blue}{[4]}}, the XYZ-correlation enables any three-party communication complexity problem to be solved with only one bit broadcast per party. Such a correlation cannot be realized via local measurements on a tripartite quantum state. This naturally raises the question of whether it can be simulated using a non-signaling quantum oracle. Since it is known that XYZ-correlation can be simulated with three PR boxes—one shared between each pair of parties \hyperlink{1}{\textcolor{blue}{[1]}}—and PR boxes themselves can be simulated with a non-signaling quantum oracle, XYZ-correlation is also simulable with such resources. However, as shown in Fig.~\ref{figS1}, a more efficient simulation protocol exists.

After the action of the oracle, the state of the three primed systems become
\begin{align}
	\ket{\psi}_{A'B'C'}=\tfrac{1}{2}\Big(\ket{00(xyz\oplus0)}+\ket{01(xyz\oplus1)}+\ket{10(xyz\oplus1)}+\ket{11(xyz\oplus0)}\Big)_{A'B'C'}. 
\end{align}
This oracle acts as a NS resource when their inputs are limited to computational states. However, this limited access oracle simulates the XYZ-correlation if Alice, Bob, and Charlie performs computational basis measurement on their respective primed qubit. It this quantum oracle is replaced by its classical counterpart, the sate of the primed systems after the oracle action reads as 
\begin{align}
	\rho_{A'B'C'}&=\tfrac{1}{2}\Big((0)_{A'}(0)_{B'}(xyz\oplus0)_{C'}+(0)_{A'}(1)_{B'}(xyz\oplus1)_{C'}+(1)_{A'}(0)_{B'}(xyz\oplus1)_{C'}\nonumber\\
	&\hspace{8cm}+(1)_{A'}(1)_{B'}(xyz\oplus0)_{C'}\Big), 
\end{align}
which also perfectly simulates the XYZ-correlation. This particular correlation can be generalized for arbitrary $n$-number of parties, and the resulting correlation reads as 
\begin{align}
	p(a_1,a_2,\cdots,a_n|x_1,x_2,\cdots,x_n)=\begin{cases}
		\tfrac{1}{2^{n-1}},~~~a_1\oplus a_2\oplus\cdots\oplus a_n=x_1x_2\cdots x_n;\\
		0,~~~\text{otherwise}.  
	\end{cases}
\end{align}
It is not hard so see that the aforementioned simulation protocols (quantum as well as the classical) generalize to the arbitrary many-party scenario. 

\hrule

\vspace{0.5 cm}
\noindent \hypertarget{1}{[1]} J. Barrett, N. Linden, S. Massar, S. Pironio, S. Popescu, and D. Roberts. Nonlocal correlations as an information-theoretic
resource. \href{https://journals.aps.org/pra/abstract/10.1103/PhysRevA.71.022101}{Phys. Rev. A, 71:022101 (2025)}\\

\noindent \hypertarget{2}{[2]} S. Pironio, J-D Bancal, and V. Scarani. Extremal correlations of the tripartite no-signaling polytope. \href{https://iopscience.iop.org/article/10.1088/1751-8113/44/6/065303}{J. Phys. A: Math.
	Theo., 44(6):065303 (2011).}\\

\noindent \hypertarget{3}{[3]} Ravishankar Ramanathan, Jan Tuziemski, Michal Horodecki, and Pawel Horodecki. No Quantum Realization of Extremal
No-Signaling Boxes. \href{https://journals.aps.org/prl/abstract/10.1103/PhysRevLett.117.050401}{Phys. Rev. Lett., 117:050401 (2016).}\\

\noindent \hypertarget{4}{[4]} W. van Dam. Implausible consequences of superstrong nonlocality. \href{http://dx.doi.org/10.1007/s11047-012-9353-6}{Natural Computing, 12(1):9–12 (2012).}\\

\noindent \hypertarget{5}{[5]} C. ´Sliwa. Symmetries of the bell correlation inequalities. \href{http://dx.doi.org/10.1016/S0375-9601(03)01115-0}{Phys. Lett. A, 317(3–4):165–168 (2003).}

\end{document}